\documentclass{ere04}

\usepackage{graphicx}    

\def\beq{\begin{equation}}
\def\eeq{\end{equation}}

\def\bea{\begin{eqnarray}}
\def\eea{\end{eqnarray}}
\def\ba{\begin{array}}                  
\def\ea{\end{array}}

\begin{document}

\title*{On global models for finite rotating objects in
equilibrium in cosmological backgrounds}
\titlerunning{Finite rotating objects in cosmological backgrounds}
\author{B. Nolan and R. Vera}
\institute{School of Mathematical Sciences, Dublin City University, Dublin 9,
Ireland\\
\texttt{brien.nolan@dcu.ie, raul.vera@dcu.ie}}
%
%
\maketitle

The studies in general relativity
of rotating finite objects in equilibrium have
usually focused on the case when they are truly isolated,
this is, the models to describe finite objects are embedded in an
asymptotically flat exterior vacuum.
Known results ensure the uniqueness of the vacuum exterior field
by using the boundary data for the exterior field
given at the surface of the object plus the decay of the exterior
field at infinity.
The final aim of the present work is to study the consequences on
the interior models by changing the boundary condition at infinity
to one accounting for the embedding of the object in a
cosmological background. Considering first the FLRW standard
cosmological backgrounds, we are studying the general matching of
FLRW with stationary axisymmetric spacetimes in order to
find the new boundary condition for the vacuum region. Here we
present the first results.

\section{Introduction}
\label{sec:1}
The studies in General Relativity of rotating finite objects (or
local systems), i.e. astrophysical objects as stars, planet
orbits, galaxies, clusters, etc, in equilibrium have usually
focused on the case when they are truly isolated, this is, so that
the exterior field tend to zero as moving away from the object.
For account for that, the models constructed to describe the
finite object are embedded in an {\em asymptotically flat exterior
vacuum region}.

Most of the work produced along these lines 
have followed the theoretical approach based on the construction of
global models by means of the matching of spacetimes:
the whole configuration is composed by two regions,
one spacetime describing the interior of the object and another
to describe the vacuum exterior,
which have been matched across the surface of the object at all times $\Sigma$.
To account for the equilibrium state of the rotating configuration,
the whole matched spacetime, and hence the interior
and exterior regions, are to be strictly stationary.
In addition, it has been usually naturally assumed that the model
is axially symmetric.

The stationary and axisymmetric exterior vacuum region
is described by two functions $(U,\Omega)$,
that satisfy an elliptic system of partial differential equations,
known as the Ernst equations \cite{sol}.
The boundary conditions for the problem will come determined by the object,
on its surface ($\Sigma$), plus the asymptotic flat behaviour at infinity.
The boundary data for the pair $(U,\Omega)$ on $\Sigma$, which is determined
from the matching conditions with a given interior, consists of the values
of the pair of functions on $\Sigma$, up to a constant additive factor
for $\Omega$, plus the values of their normal
derivatives to $\Sigma$. In other words, the boundary data
is given by
$\{U|_{\Sigma},\Omega|_\Sigma+c_\Omega,
\vec{n}(U)|_\Sigma,\vec{n}(\Omega)|_\Sigma\}$,
where $c_\Omega$ is an arbitrary (real) constant
and $\vec{n}$ is the vector normal to $\Sigma$.
This constitutes a set of Cauchy data for an elliptic problem,
and therefore the problem is overdetermined (although not unique because
of $c_\Omega$).

Now, using the fall off for $U$ and $\Omega$ that asymptotic
flatness requires, known results ensure the uniqueness of the
exterior field given the object \cite{MASEuni,conv}. In fact, the
necessary conditions on the Cauchy data for the existence of the
exterior field have been also found \cite{Marcexis}. This fact
determines that not every model for the interior of a finite
object can describe an isolated finite object.

Nevertheless, in a more realistic situation for any astrophysical
object, moving away from the object one should eventually reach a
large scale region which ought to be not flat but described by a
dynamical cosmological model. In the framework mentioned above,
this means that we have to change the boundary conditions at
infinity implied by the asymptotically flat behaviour, to some
others accounting for the embedding of the object into a
cosmological background. To that end, we consider a vacuum
stationary axisymmetric region (in which a compact object could
reside) matched to a cosmological background.

To begin with, the cosmological backgrounds we consider are the
Friedman-Lema\^{\i}tre- Robertson-Walker (FLRW) models.  The
results presented here basically state that first, the slicing of
the matching hypersurface at FLRW side given by constant values of
the cosmological time coincide with that of constant value of an
intrinsically defined time (see below) at the stationary side.
More importantly, the surfaces defined by the slicing must be {\em
spheres}.

\section{Summary on junction of spacetimes}
The formalism for matching two $C^2$ spacetimes\footnote{A $C^m$
spacetime is a paracompact, Hausdorff, connected $C^{m+1}$
manifold with a $C^m$ Lorentzian metric (convention
$\{-1,1,1,1\}$).} $(\mathcal{W}^\pm,g^\pm)$ with respective
boundaries $\Sigma^\pm$ of arbitrary character, even changing from
point to point, was presented in \cite{MASEhyper}.
The starting point for a further development
of the matching conditions by subdividing them into constraint and evolution
equations by using a 2+1 decomposition was introduced in \cite{MARCaxi}
(see also \cite{MarcES}). For completeness, this section is devoted for
a summary of the formalisms. We refer to
\cite{MASEhyper,MARCaxi,MarcES} for further details.

Gluing $(\mathcal{W}^+,g^+,\Sigma^+)$ to $(\mathcal{W}^-,g^-,\Sigma^-)$
across their boundaries
consists in constructing a manifold
$\mathcal{V}=\mathcal{W}^+\cup\mathcal{W}^-$ identifying both the
points and the tangent spaces of $\Sigma^+$ and $\Sigma^-$. This
is equivalent to the existence of
 an abstract three-dimensional $C^3$ manifold
$\sigma$ and two $C^3$ embeddings 
$ \Psi_\pm~~ :~~ \sigma~ \longrightarrow ~ \mathcal{W}^\pm $ can
be constructed such that $\Psi_\pm(\sigma)=\Sigma^\pm$. Points on
$\Sigma^+$ and $\Sigma^-$ are identified by the diffeomorphism
$\Psi_-\circ\Psi_+^{-1}$. We denote by $\Sigma(\subset
\mathcal{V})\equiv  \Sigma^+ = \Sigma^-$ the identified matching
hypersurface. The conditions that ensure the existence of a
continuous metric $g$ in $\mathcal{V}$, such that $g=g^\pm$ in
$\mathcal{V}\cap\mathcal{W}^\pm$ are the so-called {\it
preliminary junction conditions} and require first the equivalence
of the induced metrics on $\Sigma^\pm$, i.e. $
  \Psi^*_+(g^+)=\Psi^*_-(g^-),
$
where $\Psi^*$ denotes the pull-back of $\Psi$. Secondly, one requires
the existence of two $C^2$ vector fields $\vec l_\pm$ defined over
$\Sigma^\pm$,
transverse everywhere to $\Sigma^\pm$, with different relative orientation
($\vec l_+$ points $\mathcal{W}^+$ inwards whereas $\vec l_-$ points
$\mathcal{W}^-$ outwards) and satisfying
$
  \Psi^*_+(\mathsf{l}_+)=\Psi^*_-(\mathsf{ l}_-),
  \Psi^*_+(\mathsf{ l}_+(\vec l_+))=
  \Psi^*_-(\mathsf{ l}_-(\vec l_-)),
$
where $\mathsf{ l}_\pm=g^\pm(\vec l_\pm,\cdot)$.
The existence of these so-called rigging vector fields is not ensured
when the boundaries have null points \cite{signlong}.

Now, the Riemann tensor in $(\mathcal{V},g)$ can be
defined in a distributional form (see \cite{MASEhyper}). In order
to avoid singular terms in the Riemann tensor
on 
$\Sigma$, a second set of conditions must be imposed. This second set
demands the equality of the so-called generalized second fundamental forms
with respect of the rigging one-forms, and can be expressed as
\begin{equation}
  \label{eq:second}
  \Psi^*_+(\nabla^+\mathsf{ l}_+)=\Psi^*_-(\nabla^-\mathsf{ l}_-),
\end{equation}
where $\nabla^\pm$ denotes the Levi-Civita
covariant derivative in $(\mathcal{W}^\pm,g^\pm)$.
If (\ref{eq:second}) are satisfied for one choice
of pair of riggings, then they do not depend on the choice of riggings
\cite{MASEhyper}.

Once the whole set of matching conditions hold, the finite one-side limits of
the Riemann tensor of $(\mathcal{V},g)$ on $\Sigma$,
and in any $C^1$ coordinate system covering $\Sigma$ (or part thereof),
satisfy the following relation
\begin{equation}
  \label{eq:riemanns}
  \left.R^+_{\alpha\beta\mu\nu}\right|_{\Sigma}=\left.
  R^-_{\alpha\beta\mu\nu}+n_\alpha n_\mu B_{\beta\nu}
  -n_\beta n_\mu B_{\alpha\nu}-n_\alpha n_\nu B_{\beta\mu}+
  n_\beta n_\nu B_{\alpha\mu}\right|_{\Sigma},
\end{equation}
where $R^\pm_{\alpha\beta\mu\nu}$ are the Riemann tensors of
$(\mathcal{W}^\pm,g^\pm)$, respectively,
$\mathsf{n}$ is the normal
one-form to $\Sigma$, and $B_{\alpha\beta}$ is a symmetric tensor
which is defined up to the transformation
$
  B_{\alpha\beta}~\rightarrow~ B_{\alpha\beta}+X_\alpha n_\beta+
  X_\beta n_\alpha,
$
for arbitrary one-form $\mathsf{X}$.

Following \cite{MARCaxi,MarcES}, the 2+1 splitting of the matching
conditions starts by foliating $(\sigma,\Psi^*_-(g^-))$ with a set
of spacelike $C^3$ two-surfaces $\sigma_\tau$ where $\tau\in
\mathrm{I\hspace{-2.8pt}R}$. Let $i_\tau: \sigma_\tau\rightarrow
\sigma$ be the inclusion map of $\sigma_\tau$ into $\sigma$. The
compositions $\Psi_{\tau,\pm}\equiv \Psi_\pm\circ i_\tau$ define
embeddings of $\sigma_\tau$ into $(\mathcal{W}^\pm,g^\pm)$, and
the images $S^\pm_\tau\equiv \Psi_{\tau,\pm}(\sigma_\tau)$ are
spacelike two-surfaces lying on $\Sigma^\pm$ by construction.
Clearly, the identification of $\Sigma^+$ with $\Sigma^-$ through
$\Psi_-\circ\Psi_+^{-1}$ induces the identification of $S_\tau^+$
with $S_\tau^-$ by the diffeomorphism $\Psi_{\tau,+}\circ
({\Psi_{\tau,-}})^{-1}$. The identified surfaces will be denoted
by $S_\tau\equiv S_\tau^+=S_\tau^-$, and thus
$S_\tau\subset\Sigma$. For any given point $x\in S_\tau$, let us
denote by $N_x S^\pm_\tau$ the two-dimensional Lorentzian vector
space, subset of the cotangent space $T^*_x\mathcal{W}^\pm$,
spanned by the normal one-forms of $S^\pm_\tau$ at $x$. The
(normal) bundle with fibers $N_xS^\pm_\tau$ and base $S^\pm_\tau$
will be denoted by $NS^\pm_\tau$.

The matching conditions impose restrictions on $S_\tau$ for each value
of $\tau$. These are called the
{\it constraint matching conditions} and consist of two parts.
First, the restriction of the preliminary junction conditions on $S_\tau$
imposes the isometry of $S^+_\tau$ and $S^-_\tau$, i.e.
\begin{equation}
  \label{eq:preconstraint}
  \Psi^*_{\tau,+}(g^+)=\Psi^*_{\tau,-}(g^-).
\end{equation}
Secondly, and in order to ensure the identification of the tangent
spaces in $\Sigma^\pm$, for every $x\in S_\tau$ there must exist
a linear and isometric map
\begin{equation}
  \label{eq:isomap}
  f^x_\tau~:~ N_x S^+_\tau~\longrightarrow~ N_x S^-_\tau,
\end{equation}
with the following property, inherited by (\ref{eq:second}):
the second fundamental form of $S^+_\tau$ with respect to any
$\mathsf{n}\in NS^+_\tau$, denoted by
$\mathsf{K}^+_{S_\tau}(\mathsf{n})\equiv \Psi^*_{\tau,+}(\nabla^+\mathsf{n})$,
and the corresponding image through $f_\tau$,
i.e. the one-form field $f_\tau(\mathsf{n})$ to $S^-_\tau$,
will have to coincide, i.e.
\begin{equation}
  \label{eq:secconstraint}
  \mathsf{K}^+_{S_\tau}(\mathsf{n})=\mathsf{K}^-_{S_\tau}(f_\tau(\mathsf{n})),\hspace{1cm}
  \forall \mathsf{n}\in NS^+_\tau.
\end{equation}

For further details and more explicit forms of the above expressions
we refer to \cite{MARCaxi}.

\section{Matching FLRW with stationary and axisymmetric spacetimes}
Regarding the FLRW spacetime, and since we will follow the procedures
used in \cite{MarcES}, let us review
some notation and conventions.
\begin{definition}
Let $(\mathcal{M},g_\mathcal{M})$ be a complete, simply connected,
three-dimensional Riemannian manifold of constant curvature and
$I\subset \mathrm{I\hspace{-2.8pt}R}$ an open interval.
A FLRW spacetime $(\mathcal{V}^{RW}, g^{RW})$ is the manifold
$\mathcal{V}^{RW}=I\times \mathcal{M}$
endowed with the metric $g^{RW}=-dt^2+a^2(t)g_\mathcal{M}$, where
the so-called scale factor $a(t)$ is a positive $C^3$ function on $I$,
and such that
\begin{enumerate}
\item The energy density $\rho$ and the pressure of the cosmological
flow $p$ satisfy $\rho\geq 0$, $\rho+p\neq0$,
\item the expansion $\dot a /a $ vanishes nowhere on $I$ (dot denotes
$d/dt$).
\end{enumerate}
\end{definition}

\begin{definition}
To start with, no specific matter content in the stationary
and axisymmetric region will be assumed, although the corresponding
$G_2$ on $T_2$ (necessarily)
Abelian group \cite{carter70}
will be assumed to act orthogonally transitively (OT). Therefore,
there exist coordinates $\{T,\Phi,x^M\}$ ($M,N,...=2,3$) such that
the metric $g^{sx}$ in the OT stationary and axisymmetric region
$\mathcal{W}^{sx}$ reads \cite{sol}
\begin{equation}
  \label{eq:ds2sx}
  ds^2_{sx}=-e^{2U}\left( \mathrm{d} T+A\mathrm{d}\Phi\right)^2 +
e^{-2U}W^2 \mathrm{d} \Phi^2 + g_{MN}\mathrm{d} x^M \mathrm{d} x^N,
\end{equation}
where $U$, $A$, $W$ and $g_{MN}$ are functions of $x^M$,
the axial Killing vector field is given by $\vec\eta=\partial_{\Phi}$,
and a timelike Killing vector field is given by $\vec\xi=\partial_{T}$.
\end{definition}

Special attention is given to the one-form
$\mathsf{\zeta}\equiv -\mathrm{d} T$
and its corresponding vector field $\vec\zeta$, which
are orthogonal to the hypersurfaces of constant
$T$, 
as well as to $\vec\eta$ everywhere.
In fact, $\vec\zeta$ is intrinsically defined as the future-pointing
timelike vector field tangent to the orbits of the $G_2$ group
and also to the axial Killing vector field \cite{bardeen70}
(see \cite{sol}). Note that $\vec\zeta$ is hypersurface orthogonal,
but it is not a Killing vector field.
The norm of $\vec\zeta$ is given by
$\zeta^\alpha\zeta_\alpha=-F^2$ with
$
  F\equiv e^U(A^2W^{-2}-e^{-4U})^{1/2}.
$

We will denote by $\{\vec E_M\}$ any two linearly independent
vector fields spanning the surfaces orthogonal to the orbits of
the $G_2$ group, so that the set $\{\vec\zeta,\vec\eta,\vec E_M\}$
constitute a basis of the tangent spaces at every point in
$(\mathcal{W}^{sx},g^{sx})$. In the coordinate system used in
(\ref{eq:ds2sx}) one could simply take the choice $\vec
E_M=\partial_{x^M}$.

The only assumption made on $\Sigma$ is that of being generic,
i.e. such that the function given by the values of the
cosmological time on $\Sigma^{RW}$, say $\chi$ has no local maximum or
minimum \cite{MarcES}. The domain $\Sigma^{RW}_{0}\subset \Sigma^{RW}$
defined as those points with regular values of $\chi$
is then dense in $\Sigma^{RW}$ \cite{MarcES}.
Nevertheless, this assumption is easily
shown to be a property when the stationary region is vacuum, which
is the case in which we will be interested in eventually.

\begin{proposition}
Let $(\mathcal{V},g)$ be the matching spacetime between a FLRW region
$(\mathcal{W}^{RW}, g^{RW})$ and an OT stationary and axisymmetric region
$(\mathcal{W}^{sx}, g^{sx})$
across a connected, generic matching hypersurface $\Sigma$
preserving the symmetry.
Let $S^{RW}_\tau$ be the natural foliation in FLRW given by
$\Sigma^{RW}\cap\{t=\tau\}$.
Then, the following geometrical properties hold:
\begin{enumerate}
\item $\vec\zeta$ is orthogonal to each surface
$S_\tau$, and hence $S^{RW}_\tau$, at any point $p\in S_\tau$.
\item Each connected component of $S_\tau$, and hence
$S^{RW}_\tau$ and $S^{sx}_{\tau}$ is a two-sphere
with the standard metric and it is an umbilical submanifold
in $(\mathcal{V},g)$. Furthermore, there exists a spherically
symmetric coordinate system $\{t,r,\theta,\phi\}$ in
$(\mathcal{W}^{RW},g^{RW})$ such tht this surface corresponds to
$r=const.$ and $t=const.$
\end{enumerate}
\end{proposition}
{\it Proof:} We start by fixing a regular value $\tau_0$
of $\chi$ (see above) and the
corresponding surface $S^{RW}_{\tau_0}$.
For any point $p\in S^{RW}_{\tau_0}$ consider an open neighbourhood
$U\subset \Sigma^{RW}_0$ of $p$.
Let us denote by $\vec e_A$ ($A,B,C=1,2$) a pair of vector fields
on $U$ (restricting the size of $U$ if necessary)
which are linearly independent at every point and
tangent to the foliation $\{S^{RW}_\tau\}$,
and define $h_{AB}=g(\vec e_A,\vec e_B)|_U$,
where $g(\cdot,\cdot)$ represents the scalar product in the matched
spacetime $(\mathcal{V},g)$.
To complete the basis of $T_q \mathcal{V}$ for every $q\in U$ we take the
restriction of
fluid velocity vector on $U$, $\vec u|_U$, and the vector field $\vec s$,
defined as the unit normal vector of $S^{RW}_\tau$
which is tangent to $\{t=\tau\}$ (and points inwards in
$\mathcal{W}^{RW}$). 
The vector $\vec s$ is thus spacelike and transverse to $\Sigma^{RW}$
at non-critical values of $\chi$, i.e., it is transverse to $\Sigma^{RW}_0$
and thus $\mathsf{n}(\vec s)\neq 0$ in $U$. By construction, $\vec u|_U$
and $\vec s$ are mutually orthogonal and also orthogonal to $\vec e_A$.
By the 
identification of $\Sigma^{sx}$ and
$\Sigma^{RW}$ in $\Sigma\subset \mathcal{V}$, the vector field $\vec \zeta$
at any $q\in U$ can be expressed in the basis
$\{\vec u|_U,\vec s, \vec e_{A}\}$ as\footnote{For the sake of
simplicity in the following expressions, vectors (and functions)
$\vec v$ defined only on $U$
will appear as either $\vec v$ or the redundant expression $\vec v|_U$.}
\begin{equation}
  \label{eq:stU}
  \left.\vec\zeta\right|_U=
  \left. F \cosh\beta\, \vec u -F\sinh\beta\cos\alpha ~ \vec s
  +c^A \vec e_A\right|_U,
\end{equation}
where 
$\alpha$, $\beta$, $c^A$ are
scalar functions on $U$ and $c^A$ satisfy
$c^A c^B h_{AB}=F^2\sinh^2\beta\sin^2\alpha$.

Because of the preservation of the axial symmetry
\cite{mps},
the restriction to
$\Sigma$ of 
$\vec\eta$
will have to be tangent to $\Sigma$ and tangent
to the restriction to $\Sigma$ of a Killing vector field
in $(\mathcal{W}^{RW},g^{RW})$, say $\vec\eta_{RW}$,
which in turn will be also tangent to
the foliation $\{S_\tau\}$. This means
$
  \left.\vec\eta\right|_U=\left.\vec\eta_{RW}\right|_U=
  \left. \eta^A \vec e_A\right., 
$
for some functions $\eta^A$ defined on $U$.
The mutual orthogonality of $\vec\zeta$ and $\vec\eta$ demands that
\begin{equation}
  \label{eq:orth1}
  c^A \eta^B h_{AB}=0.
\end{equation}
It can be easily checked that the following two vector
fields defined on $U$,
\begin{equation}
  \label{eq:vs}
  \left.\vec v_A \right. 
  =\left.\left[\mathsf{n}
    \left(\vec s-\tanh\beta\cos\alpha\vec u\right)\right]\vec e_A+
  \frac{h_{AB}c^B}{F\cosh\beta}
  \left[\mathsf{n}(\vec s)\vec u -\mathsf{n}(\vec u) \vec s\right]\right|_U,
\end{equation}
are tangent to $\Sigma$ and orthogonal to $\vec\zeta$.
From
$
g(\vec v_A,\vec\eta)|_U=\left[\mathsf{n}
    \left(\vec s-\tanh\beta\cos\alpha\vec u\right)\right]|_U h_{AB} \eta^B
$
we see that the vector
$
  \vec v\equiv c^A \vec v_A
$
on $U$, apart from being tangent
to $\Sigma$ and orthogonal to $\vec\zeta|_U$ by construction,
is also orthogonal
to $\vec\eta|_U$ by virtue of (\ref{eq:orth1}).
Therefore, there exist two functions $a^M$ on $U$
such that $\vec v=a^M\vec E_M|_U$, as follows from (\ref{eq:ds2sx}).

The Riemann tensor in the FLRW region reads
\[
R^{RW}_{\alpha\beta\mu\nu}=
\frac{\varrho+p}{2}\left[
 u_\alpha u_\mu g^{RW}_{\beta\nu}
-u_\alpha u_\nu g^{RW}_{\beta\mu}
+u_\beta u_\nu g^{RW}_{\alpha\mu}
-u_\beta u_\mu g^{RW}_{\alpha\nu}
\right]
+\frac{\varrho}{3}
\left(
 g^{RW}_{\alpha\mu}g^{RW}_{\beta\nu}
-g^{RW}_{\alpha\nu}g^{RW}_{\beta\mu}
\right).
\]
Due to the orthogonal transitivity 
in the stationary and axisymmetric region,
we have
$
  R^{sx}_{\alpha\beta\mu\nu} \zeta^\alpha \eta^\beta \eta^\mu E^\nu_M=0
$
for $M=2,3$.
As a result, the contraction of (\ref{eq:riemanns})
with $\zeta^\alpha|_U$, $\eta^\beta|_U$, $\eta^\mu|_U$ and $v^\nu$ 
leads to
$
  0=-\left.(\varrho+p)/2~g(\vec u,\vec \zeta)~ g(\vec u,\vec v_C)~
  h_{AB}\eta^A \eta^B c^C\right|_U,
$
which by virtue of (\ref{eq:vs}) and (\ref{eq:stU}) can be expressed as
\begin{equation}
  \label{eq:main1}
  0=-\left.\frac{\varrho+p}{2}~\mathsf{n}(\vec s) h_{CD} c^C c^D ~
    h_{AB} \eta^A \eta^B\right|_U.
\end{equation}
Using $\varrho+p\neq 0$, the fact that $h_{AB}$ is positive definite
and that $\vec \eta|_U(=\eta^A \vec e_A)$
only vanishes at points in the axis,
one has that $c^A$ vanish on a dense subset of
$U$, hence $c^A=0$ for $A=1,2$ by continuity.
Therefore (\ref{eq:stU}) becomes
\begin{equation}
  \label{eq:stUf}
  \left.\vec\zeta\right|_U=\left.
F\cosh\beta\, \vec u - F\sinh\beta~ \vec s \right|_U,
\end{equation}
with a change in sign in $\beta$ if necesary,
and conclusion (1) follows.

Expression (\ref{eq:stUf}) implies $\mathsf{\zeta}|_q\in N_qS^{sx}_\tau$
for every $q\in U$, and it can be
reexpressed by (\ref{eq:isomap}), using $g^{RW}$ to lower the indices
of $\vec u$ and $\vec s$,
as
\begin{equation}
  \label{eq:fst}
  f^q_\tau(\mathsf{\zeta}|_q)=\left.F \cosh\beta~ \mathsf{u} -
    F\sinh\beta~ \mathsf{s}\right|_q
\end{equation}
for every $q\in U$.
It is now convenient to introduce the vector field $\vec\lambda$
on $U$ defined as
$
\vec\lambda=\epsilon^{AB}h_{BC}\eta^C \vec e_A\equiv \lambda^A\vec e_A,
$
where $\epsilon^{AB}=-\epsilon^{BA}$, $\epsilon^{12}=1$,
which is tangent to the foliation $\{S^{RW}_\tau\}$ and
orthogonal to $\vec\eta|_U$. Since it is also orthogonal to $\vec\zeta|_U$,
then it will have the form $\vec\lambda=\lambda_{sx}^M \vec E_M|_U$
as seen from $\Sigma^{sx}$.
The components of the second fundamental form of $S^{sx}_{\tau_0}$ with
respect to $\mathsf{\zeta}$, 
$\mathsf{K}^{sx}_{S_{\tau_0}}\left(\mathsf{\zeta}|_{S^{sx}_{\tau_0}}\right){}_{AB}$,
which is symmetric,
can be computed and used to obtain
\begin{eqnarray}
  \label{eq:kst}
  &&
  \mathsf{K}^{sx}_{S_{\tau_0}}\left(\mathsf{\zeta}|_{\Psi^{sx}_{\tau_0}(x)}\right)_{AB}
  \lambda^A \lambda^B|_x=
  \mathsf{K}^{sx}_{S_{\tau_0}}\left(\mathsf{\zeta}|_{\Psi^{sx}_{\tau_0}(x)}\right)_{AB}
  \eta^A \eta^B|_x=0,\nonumber\\
  &&\mathsf{K}^{sx}_{S_{\tau_0}}\left(\mathsf{\zeta}|_{\Psi^{sx}_{\tau_0}(x)}
  \right)_{AB}\eta^A \lambda^B|_x=
  \left.\frac{\left[g(\vec\eta,\vec\eta)\right]^2}{2 W^2}
    \lambda_{sx}^M
  E_M^\alpha\partial_\alpha
  \left(\frac{e^{2U} A}{g(\vec\eta,\vec\eta)}\right)
\right|_{\Psi^{RW}_{\tau_0}(x)},
\end{eqnarray}
for every $x\in S_{\tau_0}$.
On the other hand, for $\mathsf{u}$ one gets 
\[
\mathsf{K}^{RW}_{S_{\tau_0}}\left(\mathsf{u}|_{\Psi^{RW}_{\tau_0}(x)}\right)_{AB}=
\left.-\frac{\dot a}{a} h_{AB}\right|_x.
\]
Regarding $\mathsf{s}$, the crucial point here is that, because
of the preservation of one isometry across $\Sigma$,
the second fundamental
form $\mathsf{K}^{RW}_{S_{\tau_0}}\left(\mathsf{s}|_{\Psi^{RW}_{\tau_0}(x)}\right)$
is diagonal in the basis
$\{\vec\lambda|_{S^{RW}_{\tau_0}},\vec\eta|_{S^{RW}_{\tau_0}}\}$.
Indeed, since $\{\vec \lambda|_{S^{RW}_{\tau_0}},
\vec \eta|_{S^{RW}_{\tau_0}}\}$ span the surfaces $S^{RW}_{\tau_0}$,
orthogonal to $\vec s|_{S^{RW}_{\tau_0}}$,
and hence
$g^{RW}(\vec s,[\vec\lambda,\vec\eta])|_{S^{RW}_{\tau_0}}=0$,
and due to the fact that $\vec\eta_{RW}$ is a Killing vector field
in $(\mathcal{W}^{RW},g^{RW})$, one also has
$g^{RW}(\vec \lambda,[\vec s,\vec\eta_{RW}])|_{S^{RW}_{\tau_0}}=0$.
Therefore, the following chain of identities hold:
\begin{eqnarray}
&&\mathsf{K}^{RW}_{S_{\tau_0}}\left(\mathsf{s}|_{\Psi^{RW}_{\tau_0}(x)}\right)_{AB}
\lambda^A \eta^B|_x=
\left.\lambda^\alpha \eta_{RW}^\beta\nabla^{RW}_\alpha s_\beta
\right|_{\Psi^{RW}_{\tau_0}(x)}
=\nonumber\\
&&=\frac{1}{2}\left.\left(\eta_{RW}^\alpha s^\beta\nabla^{RW}_\alpha\lambda_\beta
-\lambda^\alpha s_\beta \nabla^{RW}_\alpha \eta_{RW}^\beta
\right)\right|_{\Psi^{RW}_{\tau_0}(x)}=
\left.
\frac{1}{2} \lambda^\alpha [\vec s,\vec\eta_{RW}]_\alpha
\right|_{\Psi^{RW}_{\tau_0}(x)}=0.
\end{eqnarray}

We are now ready to apply the constraint matching equations
(\ref{eq:secconstraint}) to $\mathsf{\zeta}|_{S^{RW}_{\tau_0}}$
using (\ref{eq:fst}).
The non diagonal part of (\ref{eq:secconstraint}),
as follows from the above, leads to the vanishing
of
(\ref{eq:kst}) for every $x\in S^{RW}_{\tau_0}$,
which implies
$
\left.
  \lambda^\alpha \partial_\alpha
  \left(e^{2U} A/g(\vec\eta,\vec\eta)\right)
\right|_{S^{sx}_{\tau_0}}=0
$.
 In short, the constraint matching conditions lead us to
 $
   \mathsf{K}^{sx}_{S_{\tau_0}}\left(\mathsf{\zeta}|_{S^{RW}_{\tau_0}}\right)=0.
 $
By virtue of (\ref{eq:fst})
one finally has 
$\mathsf{K}^{RW}_{S_{\tau_0}}\left(\mathsf{u}|_{S^{RW}_{\tau_0}}\right)-
\tanh\beta \mathsf{K}^{RW}_{S_{\tau_0}}\left(\mathsf{s}|_{S^{RW}_{\tau_0}}\right)=0$.
Now, since $\dot a$ is nowhere zero by assumption, $\beta$ is nowhere
zero on $S^{RW}_{\tau_0}$, and therefore
\begin{equation}
  \label{eq:ks}
  \mathsf{K}^{sx}_{S_{\tau_0}}\left(\mathsf{m}|_{S^{RW}_{\tau_0}}\right)=
  \frac{\dot a}{a} ~
  \mathsf{m}\left(\vec u - \coth \beta~ \vec s\right)|_{S^{RW}_{\tau_0}}~
  h|_{\tau_0}
\end{equation}
for every normal one-form $\mathsf{m}$ 
to $S^{RW}_{\tau_0}$,
where $h|_{\tau_0}$ is the induced metric
on $S^{RW}_{\tau_0}$. 
Equation (\ref{eq:ks}), in particular, tells us that $S^{RW}_{\tau_0}$ is
umbilical in FLRW, and hence in the resulting matched spacetime
$(\mathcal{V},g)$.
At this point, the rest of the proof showing points (2)
and (3) follows strictly the proof of Proposition 1
and its Corollary 1 in \cite{MarcES},

\section*{Acknowledgements}
We acknowledge the algebraic manipulation computer program
CLASSI built on SHEEP for some of the calculations.
RV thanks the Irish Research Council for Science, Engineering
and Technology (IRCSET), postdoctoral fellowship Ref. PD/2002/108.

%
%
%

\end{document}